\documentclass{elsart}

\usepackage{amssymb}
\usepackage{epsfig}

\begin{document}

\begin{frontmatter}



\title{Analytical approach of late-time evolution in a torsion
cosmology}


\author{Xi-chen Ao}
\author{Xin-zhou Li\corauthref{cor}}
\corauth[cor]{Corresponding author.}
\ead{kychz@shnu.edu.cn}
\author{Ping Xi}
\ead{xiping@shnu.edu.cn}

\address{Shanghai United Center for Astrophysics(SUCA),
Shanghai Normal University, 100 Guilin Road, Shanghai 200234,China}

\begin{abstract}
In this letter, we study the late-time evolution of a torsion
cosmology only with the spin-$0^+$ mode. We find three kinds of
analytical solutions with a constant affine scalar curvature. In the
first case, it is not physical because the matter density will be
negative. In the second case, it shows that the dark energy can be
mimicked in the torsion cosmological model. In the third case, the
characteristic of late-time evolution is similar to that of the
universe of matter dominant. And we also find a kind of expression
with the non-constant curvature that the periodic character of
numerical calculation is only the reflection of solution in a
specific period of evolution. Using these expressions, we shall be
able to predict the evolution over the late-time. From this
prediction, we know the fate of universe that the universe would
expand forever, slowly asymtotically to a halt.
\end{abstract}


\end{frontmatter}

\section{Introduction}
The cosmological revolutions of the past quarter century have
changed everything about our understanding of the future of
universe. The current observations, such as SNeIa (Supernovae type
Ia), CMB (Cosmic Microwave Background) and large scale structure,
converge on the fact that a spatially homogeneous and
gravitationally repulsive energy component, referred to as dark
energy, accounts for about $70$ \% of the energy density of
universe. Some heuristic models that roughly describe the observable
consequences of dark energy were proposed in recent years
\cite{Peebles}. Dark energy can even behave as a phantom which
effectively violates the weak energy condition \cite{Caldwell}. When
the parameter of equation of state $w$ is a constant less than $-1$,
the universe ends up with a big rip singularity which is
characterized by the divergence of curvature of the universe after a
finite interval of time \cite{Kamionsowski}. For the quintessence
\cite{Hao} and phantom \cite{Li-1} models, the future course of
evolution is shown to critically depend on the potential. Specific
phantom field models may be proposed to avoid the cosmic doomsday of
big rip \cite{Hao-1}.

On the other hand, it is inspiring that one can replace physical
field by a geometry quantity in the dark energy model. The PGT
(Poincar\'{e} Gauge Theory of gravity) has \textit{a priori}
independent of local rotation and translation potentials, which
correspond to the metric-compatible connection 1-form
$\Gamma^{\mu\nu} = \Gamma^{[\mu\nu]}_adx^{a}$ and orthonormal
coframe $\vartheta^{\mu} = e^{\mu}_{a}dx^a$, where the metric is $g
=
-\vartheta^0\otimes\vartheta^0+\delta_{ij}\vartheta^i\otimes\vartheta^j$
and $\mu$, $\nu$, $\rho\cdots$ are $4d$ coordinate (holonomic)
indices and $i$, $j$, $k\cdots$ are $3d$. The gauge vector potential
associates with field strength. In our case, they are curvature and
torsion. PGT has been regarded as an interesting alternative of GR
(general relativity) because of its gauge structure and geometric
properties \cite{Ni}. The bouncing cosmological model with torsion
was suggested in Ref. \cite{Kerlick}, but the torsion was imagined
as playing role only at high densities in the early universe.
Goenner \textit{et al.} made a general survey of the torsion
cosmology \cite{Goenner}, in which the equations for all the PGT
cases were discussed although they only solved in detail a few
particular cases. Recently some authors have begun to investigate
torsion as a possible reason of the accelerating universe
\cite{Mielke}.

There are six possible dynamic connection modes \cite{Sezgin},
carrying certain spins and parity: $2^\pm$, $1^\pm$, $0^\pm$. Some
investigations showed that $0^\pm$ may well be the only acceptable
dynamic PGT torsion modes \cite{Yo}. The pseudoscalar mode $0^-$ is
naturally driven by the intrinsic spin of elementary fermions,
therefore it naturally interacts with such sources. Consequently, it
is generally thought that axial torsion must be small and have small
effects at the late time of cosmological evolution. This is a major
reason why one does not focus on this mode at the late time. On the
other hand, the scalar mode $0^+$ does not interact in any direct
obvious fashion with any known type of matter \cite{Shapiro},
therefore one can imagine it as having significant magnitude and yet
not being conspicuously noticed. Furthermore, there is a critical
non-zero value for the affine scalar curvature since $0^+$ mode can
interact indirectly through the non-linear equation. Nester and
collaborators \cite{Nester} consider an accounting for the
accelerated universe in term of the dynamic scalar torsion. In Ref.
\cite{Shie}, it was shown that the dynamic Riemann-Cartan geometry
could contribute an oscillating aspect to the acceleration expansion
rate of the Universe. We have appeared analyzing the dynamics of
this model \cite{Li}. Applying the statefinder diagnostic to the
torsion cosmology, we find that there are some typical
characteristics \cite{Sun}. An extension model with the spin-$0^+$
and spin-$0^-$ modes was also considered, but the acceleration
mechanism is still due to the spin-$0^+$ mode \cite{Chen}.

In this paper, we study the late-time evolution of a torsion
cosmology only with spin-$0^+$ mode. In a constant affine scalar
curvature case, we find three analytical solutions: the first is not
physical, conflicting with the assumption of energy positivity; the
second shows that the dark energy can be mimicked in the torsion
cosmological model; the third shows that the behavior of late-time
evolution is analogous to that of the universe of matter dominant.
To satisfy the energy positivity requirement \cite{Yo}, there is
only a critical point $(0, 0, 0)$ for the nonlinear system which is
an asymptotically stable focus in the phase space ($H$, $\Phi$, $R$)
\cite{Li}. In non-constant curvature case, we also find a kind of
expression showing that the periodic character of numerical
calculation is only the reflection of solution in a specific period
of evolution. Using these expressions, we shall be able to predict
the evolution over the late-time. From the prediction, we know the
fate of universe that the universe would expand forever, slowly
asymtotically to a halt.

\section{The model}
Nester and collaborators \cite{Nester} consider an accounting for
the accelerated universe in terms of a Riemann-Cartan geometry:
dynamic scalar torsion. The torsion and curvature 2-forms are
defined by
\begin{eqnarray}
T^{\mu}&\equiv&
\frac{1}{2}T^{\mu}_{\,\alpha\beta}\vartheta^{\alpha}\wedge\vartheta^{\beta}=d
\vartheta^{\mu}+\Gamma^{\mu}_{\nu} \wedge \vartheta^{\nu}\\
R^{\mu\nu}&\equiv&\frac{1}{2}R^{\mu\nu}_{\
\alpha\beta}\vartheta^{\alpha}\wedge\vartheta^{\beta}=d
\Gamma^{\mu\nu}+\Gamma^{\mu}_{\rho}\wedge \Gamma ^{\rho\nu}
\end{eqnarray}
which satisfy the Bianchi identities, respectively,
\begin{eqnarray}
\mathrm{D}T^{\mu}\equiv R^{\mu}_{\, \nu}\wedge \vartheta^{\nu},\quad
\mathrm{D} R^{\mu}_{\, \nu}\equiv0
\end{eqnarray}
Theoretical analysis of PGT led us to consider tendentiously two
spin-0 modes. In this case, the gravitational Lagrangian density is
\begin{equation}
\mathcal{L}[\vartheta,
\Gamma]=\frac{1}{2\kappa}[-a_{0}R+\sum^{3}_{n=1}a_{n}{\buildrel
(n)\over T}{}^{2}+\frac{b^{+}}{12}R^{2}+\frac{b^{-}}{12}E^{2}]
\end{equation}
where ${\buildrel (n)\over T}$ is the algebraically irreducible
parts of the torsion, $R$ is the scalar curvature and $E$ is the
pseudoscalar curvature \cite{Chen}. Note that $a_{0}$ and $a_{n}$
are dimensionless parameters, and $b^{\pm}$ have the same dimension
with $R^{-1}$.

Since current observations favor a flat universe, we will work in
the spatially flat Robertson-Walker cosmological model. The
isotropic orthonormal coframe has the form:
\begin{eqnarray}
\vartheta ^{0}=\d t, \qquad \vartheta^{i}=a(t)\d x^{i}
\end{eqnarray}
Because of isotropy, the only non-vanishing torsion tensor
components are
\begin{eqnarray}
T^{i}_{j0}=-\frac{\Phi(t)}{3}\delta^{i}_{j}, \qquad
T^{i}_{jk}=-2\chi(t)\epsilon^{i}_{jk}
\end{eqnarray}
where $\epsilon_{ijk}=\epsilon_{0ijk}$ is the usual asymmetric
tensor. Using the equation obtained by the variation with respect to
the connection, we know the $0^{-}$ part couples to the axial spin
vector of spin-$\frac{1}{2}$ fermions, but $0^{+}$ mode does not
couple to any known source\cite{Chen}. Therefore, one can consider
only spin-$0^{+}$ mode. In other words, we only discuss the case of
$\chi (t) = 0$. Furthermore, from the field equation one can finally
give the necessary equations for the matter-dominated era to
integrate (for a detailed discussion, see Ref.\cite{Chen} and we
have made the replacement $A_{0}\rightarrow -a_{0}$ and
$A_{n}\rightarrow 2 a_{n}$, which is consistent with an earlier work
\cite{Li}.)
\begin{eqnarray}
\dot H &=& \frac{\mu}{6 a_{2}}R-\frac{\kappa\;\rho_{m}}{6 a_{2}}-2
H^{2}\\
\dot \Phi &=& \frac{a_{0}}{2a_{2}}R-\frac{\kappa\;\rho_{m}}{2
a_{2}}-3H\Phi +\frac{1}{3}\Phi^{2}\\
\dot R &=& -\frac{2}{3}\left(R+\frac{6\mu}{b}\right)\Phi
\end{eqnarray}
where $b\equiv b^{+},\; \mu=a_{2}-a_{0}$, $H=\dot a/a$ is Hubble
parameter, and the energy density of matter component is
\begin{equation}
\kappa
\rho_{m}=\frac{b}{18}(R+\frac{6\mu}{b})(3H-\Phi)^{2}-\frac{b}{24}R^{2}-3a_{2}H^{2}
\end{equation}
The Newtonian limit requires $a_{0}=-1$.

\section{The solutions of constant scalar curvature}
From Eq. (9), it is easy to find the scalar affine curvature remains
a constant $R = -6\mu/b$ forever as long as its initial data has
this special value \cite{Nester}. In this case, Eq. (7) can be
rewritten as
\begin{equation}
\dot{H} = -\frac{3}{4}\frac{\mu^2}{a_2b}-\frac{3}{2}H^2
\end{equation}
The positivity of the kinetic energy requires $a_2 > 0$ and $b > 0$
[14], so we have the solution
\begin{equation}
H(t) = \zeta\tan[\frac{3\zeta}{2}(t_0-t)+\arctan(\frac{H_0}{\zeta})]
\end{equation}
where $\zeta = \mu/\sqrt{2a_2b}$ and $H_0 = H(t_0)$. However, such a
choice conflicts with the assumption of energy positivity in the $R
= -6\mu/b$ case.

If we audaciously relax the parameter requirement for positive
kinetic energy, \textit{i.e.}, $a_2 < -1$ and $\mu < 0$, this
phantom scenario will turn out to be interesting. Now we have the
solution
\begin{equation}
H(t) =
\frac{\xi(\xi+H_0)(\xi-H_0)^{-1}\exp{[3\xi(t-t_0)]-1}}{(\xi+H_0)(\xi-H_0)^{-1}\exp{[3\xi(t-t_0)]+1}}
\end{equation}
where $\xi = \mu/\sqrt{-2a_2b}$. When $t$ tends to infinity, $H(t)
\rightarrow \xi$, so that the dark energy can be mimicked in the
torsion cosmological model with a constant affine scalar curvature.
Using the dynamical analysis, we have pointed out that there is a
late-time de Sitter attractor \cite{Li}. Note that the solution (13)
is just corresponding to the de Sitter attractor.

Especially, as $a_2 = -1$, we have a solution
\begin{eqnarray}
H &=& \frac{2H_0}{2+3H_0t}\\
\Phi &=&
\frac{3H_0[2+(2+2H_0t)^{\frac{1}{3}}\eta]}{(2+3H_0t)[1+(2+3H_0t)^{\frac{1}{3}}\eta]}\\
R &=& 0
\end{eqnarray}
where
\begin{equation}
\eta = \frac{2^{2/3}(-3H_0+\Phi_0)}{3H_0-2\Phi_0}
\end{equation}
From Eq. (14), we have
\begin{equation}
a = a_0(\frac{2+3H_0t}{2+3H_0t_0})^{\frac{2}{3}}
\end{equation}
Note that the behavior of the late-time evolution is analogous to
that of the universe of matter dominant. In Fig. 1, we plot the
rajectories in the phase space H-$\Phi$ when we take $a_2 \leq -1$,
$b
> 0$ and $R = -6\mu/b$. Obviously, if $a_2 = 0$ or $-1$, the
slightest change in $a_2$ leads to a radical change in the behavior
of solutions. Therefore, we have a bifurcation at $a_2 = 0$ or $-1$
in Eqs. (7)-(8) and $R = -6\mu/b$.
\begin{figure}
\epsfig{file=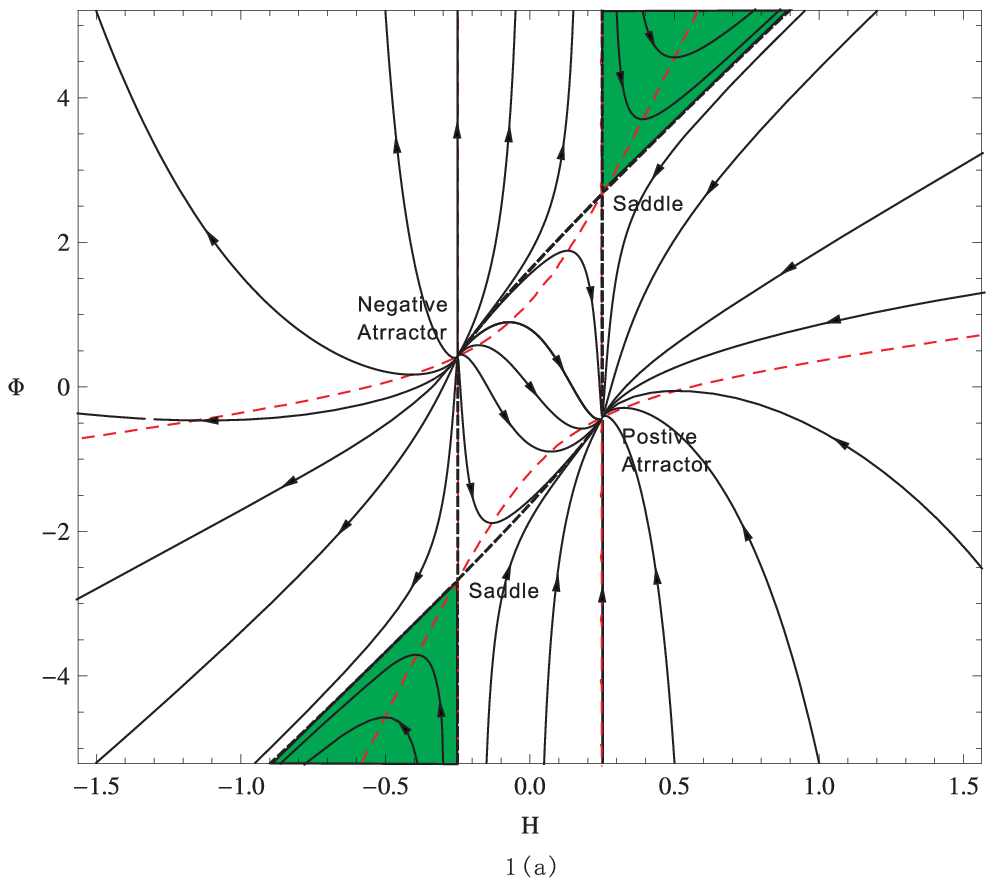,height=2.1in,width=2.5in}
\end{figure}
\begin{figure}
\epsfig{file=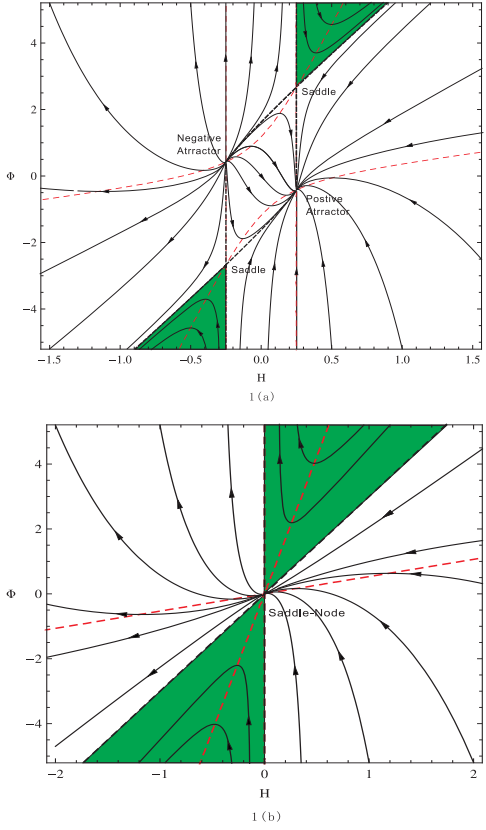,height=2.1in,width=2.5in}\caption{The
rajectories in the phase space $H$-$\Phi$ fixing $a_2 < -1$ in 1(a)
and $a_2 = -1$ in 1(b), respectively, in $R = -6\mu/b$ case.}
\end{figure}

\section{The solution of non-constant scalar curvature}
The numerical analyses show that $H$, $\Phi$ and $R$ have a periodic
character at late-time of the evolution for $a_{2}>0$ and $b>0$,
approximatively \cite{Shie}. Using the dynamical analysis \cite{Li}
and statefinder diagnostic \cite{Sun}, we find that this character
is corresponding to an asymptotically stable focus. In Fig. 2, we
plot evolving trajectory in $H$-$\Phi$-$R$ space, where we have
chosen $a_{2} = 2$ and $b = 2/t_{0}^2$. We find easily that the
evolving trajectories tend to the focus (0,0,0).
\begin{figure}
\epsfig{file=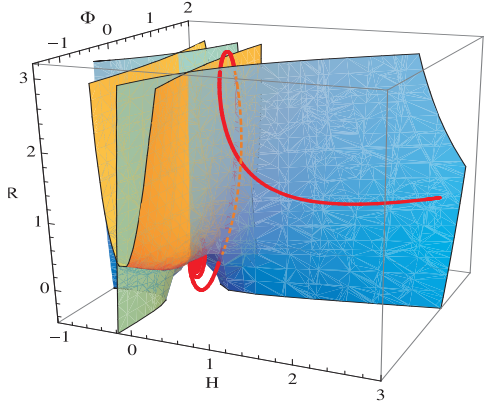,height=2.1in,width=2.5in}\caption{We
illustrate evolution trajectory in the phase space, where have
chosen $a_{2}=2, b=2/t_{0}^{2}$.  The surfaces are the nullclines of
nonlinear dynamical system, and the curve is the evolution
trajectory. It shows clearly that at late time the phase line has a
oscillation feature.}
\end{figure}
If we consider the linearized equations, then Eqs.(7)-(9) can be
reduced to
\begin{eqnarray}
\dot{H}=\frac{\mu}{6 a_{2}}R,\quad \dot\Phi=\frac{1}{2a_2}R,\quad
\dot R=-\frac{4\mu}{b}\Phi \label{linearized}
\end{eqnarray}
The linearized system (\ref{linearized}) has an exact periodic
solution
\begin{eqnarray}
H&=&-\alpha R_{0}\sin \omega t+\frac{\mu}{3}\Phi_{0}\cos \omega t
+H_{0}-\frac{\mu}{3}\Phi_{0} \nonumber \\
\Phi&=&-\beta^{-1}R_{0}\sin \omega t +\Phi_{0}\cos \omega t \nonumber \\
R&=&R_{0}\cos \omega t +\beta \Phi_{0}\sin \omega t
\end{eqnarray}
where $\omega =\sqrt{\frac{2\mu}{a_{2}b}},\  \alpha=\sqrt{\frac{b
\mu}{72 a_{2}}},\ \beta=\sqrt{\frac{8\mu a_{2}}{b}}$ and
$H_{0}=H(0),\ \Phi_{0}=\Phi(0)$ and $R_{0}=R(0)$ are initial values.
Obviously, $(H,0,0)$ is a critical line of center for the linearized
solution. In other words, there are only exact periodic solutions
for the linearized system, but there are quasi-periodic solutions
near the focus for the coupled nonlinear equations. This property of
quasi-periodic also appears in the statfinder diagnostic with the
case of $a_{2}\ge 0$ \cite{Sun}. According to nonlinear equations
(7)-(9), we can obtain the critical points and study the stability
of these points. To study the stability of the critical point $(0,
0, 0)$, we write the variables near $(0, 0, 0)$ in the form $H =
\triangle H$, $\Phi = \triangle \Phi$ and $R = \triangle R$, where
$\triangle H$, $\triangle \Phi$ and $\triangle R$ are the
perturbations of the variables near the critical point $(0, 0, 0)$.
Substituting the expressions into Eqs. (7)-(9), we obtain the
corresponding eigenvalues $(0, -\sqrt{-\frac{2\mu}{a_2b}},
\sqrt{\frac{2\mu}{a_2b}})$. Therefore, the critical point $(0, 0,
0)$ is a stable focus when $a_2 > 0$ and $b > 0$. We illustrate the
evolution trajectory in the phase space $(H, \Phi, R)$ in Fig.2,
where have chosen $a_2 = 2, b = 2/t_0^2$. Nullcline surfaces in Fig.
2 divided the phase-space $(H, \Phi, R)$ into some domains, in each
of which there is definite behavior of the evolution trajectory. It
is obvious that the evolution trajectory helically tends to the
critical point $(0, 0, 0)$ after crossing the nullcline surface.

For the late-time behavior of non-linear system Eqs.(7)-(9), we
should determine the behavior at infinity. We substitute $\tau =
\frac{1}{t}$ in Eqs. (7)-(9), and show the Laurent expansions around
the point $\tau = 0$. Clearly, as $t \rightarrow \infty$, $\tau
\rightarrow 0$. Thus, we are interested in the behavior of equations
for $\tau = 0$. The non-linear system clearly shows that the
solutions have regular behavior at $\tau = 0$, and have the Taylor
expansions: $H(t) = \sum^{\infty}_{n=1}h_n\tau^{n}, \Phi(t) =
\sum^{\infty}_{n=3}\phi_n\tau^{n}$ and $R(t) =
\sum^{\infty}_{n=2}r_n\tau^n$. So we find an approximate formula up
to $t^{-N}$ order as follows
\begin{eqnarray}
H(t)&=&\sum^{N}_{n=1}\frac{h_{n}}{t^{n}}+\frac{1}{t^{N}}[\beta_{N}\sin
\omega t +\gamma _{N} \cos \omega t]\\
\Phi(t)&=&\sum^{N}_{n=3}\frac{\varphi_{n}}{t^{n}}+
\frac{1}{t^{N}}[\frac{3\beta_{N}}{\mu}\sin \omega t
+\frac{3\gamma_{N}}{\mu}\cos \omega t ]\\
R(t)&=&\sum^{N}_{n=2}\frac{r_{n}}{t^{n}}+
\frac{1}{t^{N}}[\frac{-12\gamma_{N}}{\omega b}\sin{\omega t}
+\frac{12\beta_{N}}{\omega b}\cos{\omega t}]
\end{eqnarray}
where $\omega = \sqrt{\frac{2\mu}{a_{2} \,b}}$ and $h_{n},\
\varphi_{n}$ and $r_{n}$ are undetermined coefficients. The
coefficients $\beta_{N}$ and $\gamma_{N}$ can be written as the
terms of initial values.

From a theoretical point of view, the Eqs. (21)-(23) are equivalent
to the Laurent expansions around the point $\tau = 0$ when $N
\rightarrow \infty$. From a practical point of view, the Eqs.
(21)-(23) are more effective than the expansions of minus-power
terms because of the latter only determined by the coefficients
$h_1, \cdots, h_{N-1}$; $\varphi_1, \cdots, \varphi_{N-1}$; $r_1,
\cdots, r_{N-1}$ for a definite $N$. Therefore, the former is an
approximate formula up to $t^{-N}$ order and the latter is one up to
$t^{-(N-1)}$ order. For example, we take $N=3$ and the approximate
solution up to $t^{-3}$ term is
\begin{eqnarray}
H(t)&=&\frac{2}{3t}+\frac{t_{0}^{2}}{t^{2}}(H_{0}-\frac{1}{3t_{0}}-\frac{\mu
\Phi_{0}}{3}+\frac{\mu}{9
t_{0}})\nonumber\\
&&+\frac{t_{0}^{3}}{t^{3}}[(\frac{\mu}{3}\Phi_{0}S_{0}-
\frac{\mu}{9t_{0}}S_{0}+\frac{\omega b}{12}R_{0}C_{0}-\frac{\omega
b}{9 t_{0}^{2}}C_{0})\sin{\omega t}\nonumber\\
&&+(-\frac{\omega b}{12}R_{0}S_{0}+\frac{\omega b}{9 t_{0}^{2}}
S_{0}+\frac{\mu}{3}\Phi_{0}C_{0}-\frac{\mu}{9
t_{0}}C_{0})\cos{\omega t}]\\
\Phi(t)&=&\frac{t_{0}^{3}}{t^{3}}[\frac{b}{\mu
t_{0}^{3}}+(\Phi_{0}S_{0}-\frac{1}{3 t_{0}}S_{0}+ \frac{\omega
b}{4\mu}R_{0}C_{0}-\frac{\omega b}{3\mu t_{0}^{2}})\sin{\omega
t}\nonumber \\
&&+(-\frac{\omega b}{4\mu}R_{0}S_{0}+\frac{\omega b}{3\mu
t_{0}^{2}}S_{0}+\Phi_{0}C_{0}-\frac{1}{3t_{0}}C_{0})\cos{\omega t}]\\
R(t)&=&\frac{4}{3 t^{2}}+\frac{t_{0}^{3}}{t^{3}}[\frac{4
H_{0}}{t_{0}}-\frac{4}{3t_{0}^{2}}-\frac{4\mu}{3
t_{0}}\Phi_{0}+\frac{4\mu}{9t_{0}}\nonumber\\
&&+(R_{0}S_{0}-\frac{4}{3t_{0}^{2}}S_{0}-\frac{4\mu}{\omega
b}\Phi_{0}C_{0}+\frac{4\mu}{3\omega bt_{0}}C_{0})\sin{\omega
t}\nonumber\\
&& +(\frac{4 \mu}{\omega b}\Phi_{0}S_{0}-\frac{4\mu}{3\omega
bt_{0}}S_{0}+R_{0}C_{0}-\frac{4}{3t_{0}^{2}}C_{0})\cos{\omega
t}]\label{sol3}
\end{eqnarray}
where $H_{0} = H(t_{0}),\Phi_{0} = \Phi(t_{0})$ and $R_{0} =
R(t_{0})$, and $S_{0} = \sin{\omega t_{0}}, C_{0} = \cos{\omega
t_{0}}$. Obviously, $H(t)\rightarrow 0, \Phi(t) \rightarrow 0$ and
$R(t) \rightarrow 0$, when $t$ tends to infinity. Therefore, the
fate is that the universe would expand forever, slowly asymtotically
to a halt.

Next, we give the reason why some numerical calculations look as
periodicity as mentioned in Ref.\cite{Shie}. To put it bluntly, this
is a reflection of the late-time solution at specified period of
evolution. If we take $t = t_{0}+\bigtriangleup t,\ \bigtriangleup t
\in [0,\delta ]$ and $\frac{2\pi}{\omega}\ll\delta\ll t_{0}$, Eqs.
(24)-(26) can be reduced to a solution, whose behavior looks like a
periodic solution of (20) of the linearized system. Taking the
parameter values, we plot Hubble parameter $H(t)$ at the late-times
in Fig. 3.

If we take the approximate expression of $N = 2$, when $t
> 10, 10^2$ and $2\times10^2$, the error $\Delta =
\frac{|H_{num}-H_{N = 2}|}{H_{0}} < 0.066, 0.012$ and $0.002$,
respectively. Therefore, if we want to obtain an effective
approximate expression, $N$ is required to be large enough. In Fig.
4, a direct quantitative connection is made between analytical
expressions and the numerical solutions presented by Shie et al.
\cite{Shie}. We take initial conditions $H(1) = 1, \Phi(1) = 1.4,
R(1) = 1.53, \mu = 1.09$ and $b = 1.4$ which are the same as those
in Ref. \cite{Shie}.
\begin{figure}
\epsfig{file=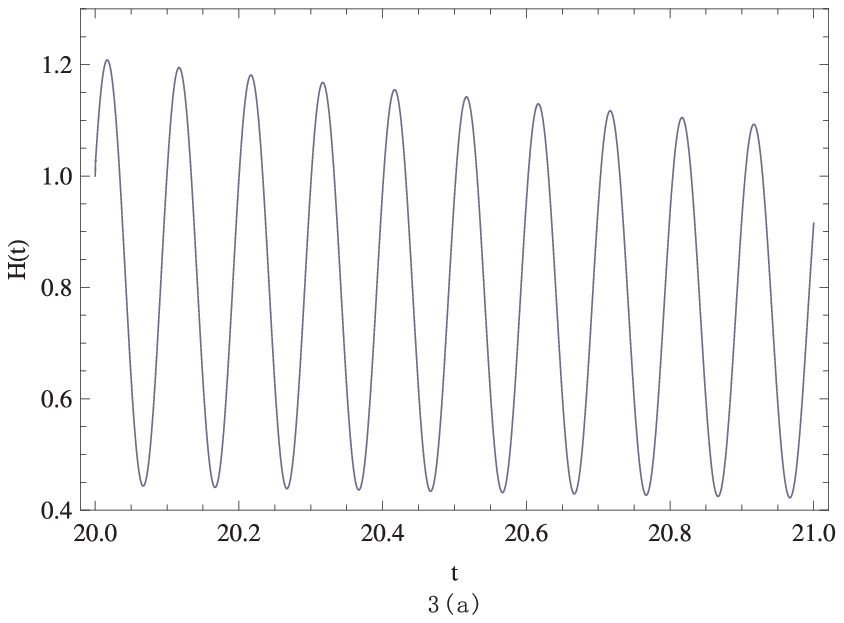,height=2.1in,width=2.5in}
\end{figure}
\begin{figure}
\epsfig{file=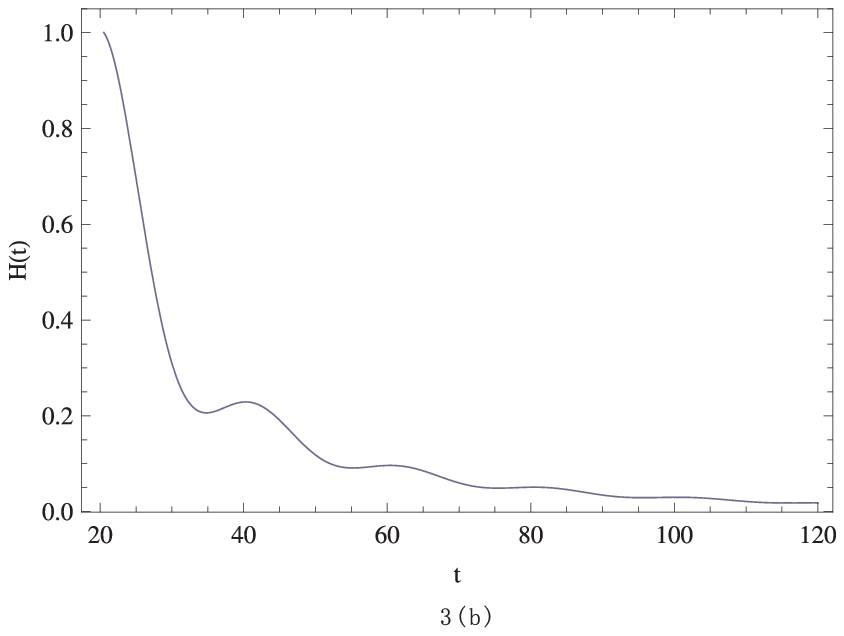,height=2.1in,width=2.5in}\caption{In the
non-constant curvature case, we plot the behavior of late-time
evolution. In 3(a), we have chosen $a_2 = 1, b =
\frac{2}{\pi^{2}t_{0}^{2}}$ and $t_0 = 20$; In 3(b), we have fixed
$a_2 = 1, b = \frac{80000}{\pi^{2}t_{0}^{2}}$ and $t_0 = 20$.}
\end{figure}
\begin{figure}
\epsfig{file=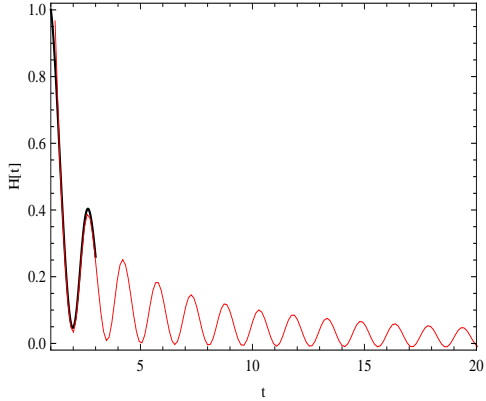,height=2.1in,width=2.5in}\caption{In the
non-constant curvature case, we plot the behavior of late-time
evolution $H(t)$ for our analytical expression and numerical
solution obtained by Shie et al., respectively. Black line
represents the numerical solution ($t \leq 3$), and red line
represents the analytical solution ($N = 50$).}
\end{figure}

\section{Conclusion}
In this paper, we study the evolution of a torsion cosmology only
with the "scalar torsion" mode $0^{+}$. This mode has some
distinctive and interesting qualities. It can be considered as a
"phantom field", because it does not interact with any known matters
directly. It only interacts via the gravitational equations.
Therefore, $0^{+}$ mode can drive the universe to accelerate at
present. However, we have to investigate the late-time solution if
we want to know the fate of universe. We find a kind of late-time
solution which  tends to the focus (0,0,0) in the phase space($H$,
$\Phi$, $R$). Under the solution , the feature of evolution is
universal with generic choices of the parameters. Distinctive and
interesting results are:
\renewcommand{\theenumi}{\roman{enumi}}
\renewcommand{\labelenumi}{\theenumi)}
\begin{enumerate}
\item In the constant affine curvature case, there are a kind
of non-physical solution and two kinds of physical solutions at
late-time. One of physical solutions corresponds to the de Sitter
attractor, and the other is analogous to the universe of matter
dominant. In the view-point of dynamical analysis, $a_2 = 0$ and
$-1$ are the bifurcations.
\item In the non-constant affine curvature case, we find a kind of
expression corresponding to the universe which would expand forever,
slowly asymptotically to a halt. The specific behaviors of late-time
evolution differ from those at different periods. If we considered
the specified period of evolution $t = t_0+\Delta t$, $\Delta t \in
[0, \delta]$, and satisfied $\frac{2\pi}{\omega} \ll \delta \ll
t_0$, the behavior looks like a periodic solution.
\end{enumerate}
Furthermore, if we want to establish a realistic cosmological model,
we have to carry out the solutions of $t < t_{0}$ period, and
compare these analytical solutions with observations to determine
the model parameters. For example, using SN Ia data, we constrain
the parameters of torsion cosmology and get their best values. Then
we can discuss the properties of the realistic cosmological models.
These issues will be considered elsewhere.

Recently, the cosmological model with even and odd parity modes in
PG has been considered by Baekler, Hehl and Nester \cite{Baekler}.
They extended the parity violating quadratic Lagrangian $V_-$ to the
general gravitational Lagrangian $V_\pm = V_- + V_+$, where $V_+$ is
the parity conserving quadratic Lagrangian. This model generalized
the torsion cosmologies which were presented by Nester et al.
\cite{Yo,Nester,Shie,Chen}. Next, we will set about investigations
for dynamics of this cosmological model.

\noindent {\bf Acknowledgments}\\
This work is supported by National Education Foundation of China
under grant No. 200931271104.

\end{document}